\documentclass[sigplan]{acmart}

\usepackage{graphicx}
\usepackage[utf8]{inputenc}
\usepackage{xspace}
\usepackage{algorithm}
\usepackage{amsmath,algpseudocode}
  \usepackage{amsthm}
\usepackage[font=small]{caption}

\newcommand{\name}{SI-HTM\xspace}
\newcommand{\powerhtm}{P8-HTM\xspace}

\newcommand{\rotbegin}{\texttt{HTMBeginROT}\xspace}
\newcommand{\rotcommit}{\texttt{HTMEnd}\xspace}

\begin{document}
%\title{Boosting Snapshot Isolation with Rollback-only Hardware Transactions}
\title[Stretching the capacity of HTM in IBM POWER]{Stretching the capacity of Hardware Transactional Memory in IBM POWER architectures}
%\title{Enhancing capacity utilization of HTM by relaxing to Snapshot Isolation }
%\numberofauthors{4}

\author{Ricardo Filipe  \hspace{0.18cm} Shady Issa \hspace{0.18cm} Paolo Romano \hspace{0.18cm} João Barreto}
\affiliation{%
  \institution{INESC-ID, Instituto Superior Técnico,  Universidade de Lisboa \\ \{ricardo.filipe, shadi.issa, paolo.romano, joao.barreto\}@tecnico.ulisboa.pt}
}
%\email{}

\begin{abstract}

%The emergence of in-memory databases and the wide adoption of weaker consistency models like snapshot isolation are prominent trends that shape the database landscape today.
%More recently, the advent of hardware transactional memory (HTM) in mainstream processors like the next ground-breaking improvement to in-memory databases. However, the limited capacity of HTM implementations is incompatible with many real-world OLTP/OLAP workloads.

The hardware transactional memory (HTM)
implementations in commercially available processors
are significantly hindered by 
%might seem like a perfect match to push the current generation of in-memory databases to unprecedented performance levels.
%However, 
their tight capacity constraints. 
In practice, this renders
current HTMs unsuitable to many real-world workloads
of in-memory databases.

This paper proposes \name, which stretches the capacity bounds of the underlying HTM, thus opening
HTM to a much broader class of applications. \name leverages the HTM implementation of the IBM POWER architecture with a software layer to offer a single-version implementation of Snapshot Isolation. 
%first solution that, to the best of our knowledge, breaks the above common belief. \name offers SI-equivalent semantics while exclusively relying on the HTM synchronization mechanisms to detect memory access conflicts and consequently abort transactions that violate SI semantics.% Therefore, \name enables state-of-the-art SI-based in-memory databases to take effective advantage of HTM.
When compared to HTM- and software-based concurrency control alternatives, \name exhibits improved scalability, achieving speedups of up to 300\% relatively to HTM on in-memory database benchmarks.

\end{abstract}




\copyrightyear{2019} 
\acmYear{2019} 
\setcopyright{acmcopyright}
\acmConference[PPoPP '19]{24th ACM SIGPLAN Symposium on Principles and Practice of Parallel Programming}{February 16--20, 2019}{Washington, DC, USA}
\acmBooktitle{24th ACM SIGPLAN Symposium on Principles and Practice of Parallel Programming (PPoPP '19), February 16--20, 2019, Washington, DC, USA}
\acmPrice{15.00}
\acmDOI{10.1145/3293883.3295714}
\acmISBN{978-1-4503-6225-2/19/02}

\maketitle

\section{Introduction}

In the quest for scalability,
in-memory databases (IMDBs) offering weak consistency guarantees
like Snapshot Isolation (SI) \cite{Berenson:1995:CAS:223784.223785}
are increasingly prominent within the database landscape.
On the one hand, the in-memory nature of IMDBs minimizes (or even eliminates) disk access latency to achieve
faster and more predictable performance \cite{7097722,Tu:2013:STM:2517349.2522713}.
On the other hand, weak consistency models alleviate many concurrency
bottlenecks that characterize serializable systems \cite{Berenson:1995:CAS:223784.223785}.
Today, popular databases like HyPer \cite{Neumann:2015:FSM:2723372.2749436}, SAP HANA \cite{DBLP:conf/icde/LeeKFMLBLLL13},
solidDB \cite{journals/debu/LindstromRRSV13} and Hekaton \cite{Diaconu:2013:HSS:2463676.2463710} combine the virtues of both trends,
relying on weakly consistent IMDB designs.

At a first glance, the recent emergence of hardware transactional memory (HTM)
support in commercially available processors such as Intel Core and IBM POWER
might seem like a perfect match to push the current generation of IMDBs to
new performance levels.
%In theory, the negligible sequential overheads of HTM could
%relieve IMDBs from the overheads of software-based lock-based
%synchronization \cite{..}.
%or impractical static partitioning approches \cite{..}.
However, the limited capacity of HTM implementations \cite{Diegues:2014:HTMLimits,Goel:2014:HTMPerf,Nakaike:2015:HTMComparison} 
is
incompatible with many real-world OLTP/OLAP workloads, whose access
footprints are often much larger than the reduced capacity of existing HTM implementations \cite{Leis2014ExploitingHT}.
%Consequently, a naive IMDB system that encapsulated each DB transaction on a HTM transaction would easily lead to prohibitive abort rates, which would easily
%degrade the performance of the system to unacceptable levels.

%. When used with appropriate workloads, such HTM implementations can achieve unprecedently low synchronization overheads with
%highly-concurrent shared-memory programs \cite{***}.

%At a first glance, HTM may seem like the next ground-breaking
%improvement to IMDBs, which would relieve them
%from costly lock-based synchronization \cite{..} or impractical
%static partitioning approches \cite{..}.

To work around this crucial obstacle, recent works either propose
modifications to HTM for SI support over a multi-versioned memory layout that eliminates the capacity limits \cite{Litz:2014:SRT:2541940.2541952,Chen:2017:AGH:3079856.3080204}; or exploit HTM as an auxiliary mechanism to accelerate software-based concurrency control schemes \cite{Leis2014ExploitingHT}.
However, while the former depends on hardware that is not yet available, the latter exploits HTM only to a limited extent.
To the best of our knowledge, the expectation of IMDBs that rely
on HTM transactions as a first-class mechanism to run and synchronize each transaction is yet to be met in practice.

%While some authors have proposed alternatives to overcome this limitation,
%such solutions incur substantial software instrumentation that blurs the pure power of HTM \cite{..}.

%the best-effort nature of mainstream HTM implementations incurs
%fundamental limitations that strongly hinder their effectiveness in the
%context of in-memory databases.

%On the other hand, existing HTM implementations are strongly consistent by nature, ensuring opacity guarantees \cite{..}.
%Hence, even in workloads where transactions are small enough to fit in HTM,
%the potential scalability is limited by the strong restrictions that opacity places on the correct parallel executions. While it is well studied that weaker consistency criteria, such as SI, can achieve higher scalability by breaking such strong restrictions, that option is simply not offered by mainstream HTM implementations.
%Hence, in-memory database designs need to decide between strongly consistent synchronization with HTM, or weak consistency software concurrency control.

%This paper questions the common belief that existing HTM implementations are
%effectively incompatible with weakly consistent in-memory databases.
This paper proposes \name, the first solution that achieves such a goal by relying on a commercially available HTM implementation -- the HTM support originally introduced in the IBM POWER8 architecture and continued in the most recent IBM POWER9 \cite{ibm_2018}. Hereafter, we will denote such HTM implementation as \powerhtm.
%Focusing on the POWER 8 HTM implementation, we propose \name, the
%first solution that, to the best of our knowledge, breaks the above
%common belief.
The key novelty of \name is that, with no hardware modification to \powerhtm,
\name is able to support SI-equivalent guarantees while relying on the hardware to detect conflicts and abort transactions.

Intuitively, \name constructs a restricted SI implementation by combining two building
blocks: i) rollback-only transactions (ROT), a complementary mode available in \powerhtm that is originally aimed at speculative execution of code blocks that do not manipulate shared data \cite{ibm_2018}; and ii) a software regulated quiescence phase that is added before the hardware commit to ensure that the transaction only commits once it is certain that its execution is compliant with SI semantics.

%In \name, memory reads and writes are executed directly on the HTM. Furthermore, \name is able to run read-only (RO) transactions as non-ROT blocks, hence avoiding any HTM-induced aborts for RO transactions.

As we describe later, this hybrid software-hardware mechanism is
able to substantially stretch the capacity bounds of the hardware transactions that can run on \powerhtm, with no software instrumentation of memory reads and writes.
%In the case of read-write transactions, 
\name eliminates capacity
bounds on a transaction's read set, restricting only their write sets
by the HTM capacity. Since many IMDB workloads are dominated by read-only and read-dominated transactions with few writes, \name is typically able to run the vast majority of transactions in the HTM fast path.

%  More precisely, \name eliminates any hardware limits on the read-set size. Hence, capacity aborts are essentially reduced to write-set capacity aborts.

%while exclusively relying on HTM transaction to completely execute each IMDB transaction.%, with no software instrumentation of the memory accesses performed by the transaction.
%More concretely, each Therefore, \name allows enables state-of-the-art SI-based in-memory databases to
%take effective advantage of HTM.

Breaking the tight capacity bounds of the original \powerhtm contributes to
important improvements in the scalability that \powerhtm can attain, %when compared to the baseline alternative of running DB transactions encapsulated on plain HTM transactions,
for two distinct reasons. First and foremost, with \name, less transactions abort due to exceeding the HTM capacity. This means less frequent situations that require falling
back to a sequential fall-back path. % global lock, during which one transaction runs sequentially and halts all the remaining ones.
The second improvement is related to the notable power of simultaneous multi-threading (SMT) on the POWER8 and 9 architectures, which are able to run up to 8 hardware threads on each core. As acknowledged by previous studies on \powerhtm~\cite{Nakaike:2015:HTMComparison}, this SMT feature is practically incompatible with HTM programs since the already scarce HTM capacity becomes shared among the co-located SMT threads. By stretching the capacity of \powerhtm, \name enables SMT-friendly transactional workloads to achieve speed-ups at multi-SMT levels, thus enabling parallelism in scenarios that are typically strongly adverse to HTM based applications.

%\item Higher scalability. This is the intuitively-expected advantage. Since \name implements SI-equivalent semantics, which are weaker than opacity, more memory access interleavings are allowed between concurrent transactions. Therefore, it is natural that transactional conflicts are less prone to occur in \name than in a pure HTM-based approach.

%\item Support for larger transactions. Perhaps surprisingly, the main advantage of \name is its ability to dramatically

%\end{itemize}

This paper has three main contributions:

\begin{itemize}

\item We propose \name, a restricted implementation of SI for \powerhtm.

\item We experimentally evaluate \name on a real IBM POWER8 server,
both with a synthetic benchmark and the TPC-C benchmark 
\cite{TPCC},
which is serializable under SI.
When compared to HTM-based concurrency control alternatives, \name exhibits speedups of up to 300\% on TPC-C.% and *** relatively to HTM and software-based concurrency control solutions, respectively.

\item We prove that any execution history that \name allows is correct under SI.
%is a correct, yet restricted single-version implementation of SI.
An important corollary is that any application that is serializable under SI is also serializable on \name.

\end{itemize}

The remainder of this paper is organized as follows. Section \ref{sec:background} provides the background on SI and on the features of \powerhtm.
Section \ref{sec:implementation} describes \name.
%Section \ref{sec:correctness} proves that \name is a correct, yet restricted, implementation of SI. 
Section \ref{sec:evaluation} evaluates \name, comparing its performance to relevant alternatives. Section \ref{sec:relwork} surveys
related work. Finally, Section \ref{sec:conclusions} concludes
and describes future work.

\section{Background}
\label{sec:background}

In this section we start by introducing the basic notions of SI. We then describe the HTM support in the IBM POWER architecture (which we
call \powerhtm), emphasizing its limitations and showing why it is not trivial to obtain weak semantics like SI when using this HTM.

%######################################################
\subsection{Snapshot Isolation}
%######################################################

SI is a widely used correctness criterion in databases.
Intuitively, SI allows each transaction to
read from and write to its own private isolated snapshot of data \cite{Berenson:1995:CAS:223784.223785,Fekete:2005:MSI:1071610.1071615,Cerone:2016:ASI:2933057.2933096}.

Each transaction's snapshot is created when the transaction starts (or, alternatively, when it performs its first read). Hence the snapshot holds the committed values that were valid at that moment.

Each transaction's snapshot is isolated from the writes of concurrent transactions.
More precisely, each write that an active transaction performs on its
snapshot is only visible to that transaction; other concurrent transactions
will not observe such write when reading from their own snapshots.
This means that SI is typically implemented in a \emph{multi-versioned} approach, since different transactions reading from the same location at the same time may observe different versions.
It is only when a transaction commits that its writes become (atomically) visible to any new transaction whose snapshot is created afterwards.

SI aborts transactions in the presence of write-write conflicts.
More precisely, a transaction $t$ is only allowed to commit if its write set
does not overlap with the write-set of any other (concurrent) transaction that
has committed after $t$ started.
%\footnote{Actually, different concrete
%proposals exist for the write-write conflict restriction, including 
%\emph{first committer wins} and \emph{first writer wins} \cite{*}.}
In contrast, SI tolerates read-write conflicts. Therefore, two transactions may commit even if one transaction's read-set overlaps with the other transaction's write-set.

\begin{figure}[t]
\centering
\includegraphics[width=0.5\textwidth]{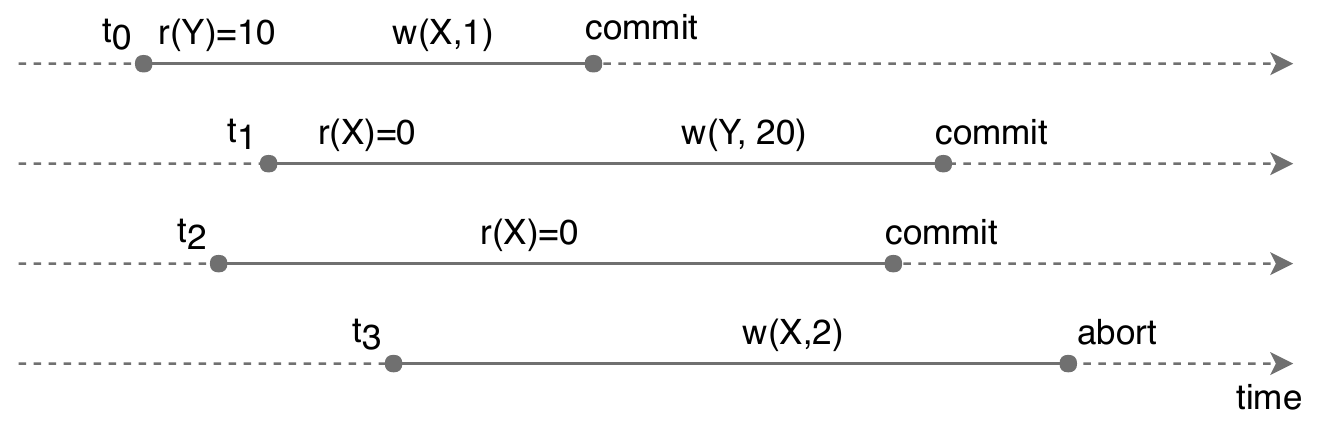}
\caption{Example of SI semantics.}
\label{fig:SI-basics}
\end{figure}

To illustrate SI semantics, consider the example in Figure \ref{fig:SI-basics}.
Since $t_1$ and $t_2$ read from their own snapshots, they are isolated from the writes that $t_0$ performs. Since $t_1$ and
$t_2$ do not incur any write-write conflict, they can safely commit under SI. Only $t_3$ has to abort, because of the write-write conflict with $t_0$ (on $X$). Note, however, that SI allows both $t_0$ and $t_1$ to commit, although they are not serializable.

%Hence, SI is weaker than serializability (or opacity \cite{Guerraoui:2008:CTM:1345206.1345233}).
The weaker (than serializable) guarantees of SI enable efficient concurrency control implementations and improved concurrency, especially for read-only or read-dominated transactions, while avoiding most isolation anomalies \cite{Berenson:1995:CAS:223784.223785}.
These advantages have quickly rendered SI a mainstream consistency guarantee
in the database domain \cite{Ports:2012:SSI:2367502.2367523} and,
more recently, in IMDBs \cite{Neumann:2015:FSM:2723372.2749436,DBLP:conf/icde/LeeKFMLBLLL13,journals/debu/LindstromRRSV13,Diaconu:2013:HSS:2463676.2463710} and distributed transactional systems \cite{Salomie:2011:DEM:1966445.1966448,Zamanian:2017:EMD:3055330.3055335}.

SI can still yield a few anomalies, most notably the write skew anomaly \cite{Fekete:2005:MSI:1071610.1071615}.
An example of a write skew is when two transactions start from a common snapshot, each one writes to a different location, and later each transaction reads from the location written by the other transaction.
%For instance, transactions $t_1$ and $t_2$ start from the same snapshot, $t_1$  writes $X=1$ while $t_2$ writes $Y=2$, and finally $t_1$ reads Y and $t_2$ reads X. Clearly, this execution is impossible in a serializable system.
While some application semantics naturally tolerate the write skew anomaly, other applications may suffer from unexpected behavior in the presence of write skews.
Fortunately, recent tools and methodologies have been proposed to detect and remove write skews \cite{Fekete:2005:MSI:1071610.1071615}, with the goal of ensuring serializable executions even when the program runs under SI.
One common fix is read promotion \cite{Fekete:2005:MSI:1071610.1071615,Litz:2014:SRT:2541940.2541952}: the problematic reads are also inserted into the transaction's write set, which ensures that a write skew triggers an abort.

%######################################################
\subsection{\powerhtm}
%######################################################

%joao: capacity buffers ou transactional buffers?
%joao: rever dimensao citada abaixo para P9
\powerhtm detects conflicts by adopting a 2-Phase Locking (2PL) scheme at the granularity of a cache line. 
In \powerhtm's 2PL scheme, the last transaction to read onto some shared variable will kill the execution of any other previous writer transaction on that same variable. In the case of write-write conflicts the last writer is killed. %Since ROTs do not track reads, only read-write transactions can be aborted in this mode. Furthermore, read-write conflicts are only triggered after some transaction has written onto a shared memory location. This conflict detection scheme does not directly translate to any known memory consistency model.

\powerhtm can only handle transactions whose read and write sets fit into each core's transactional buffer. For the IBM POWER8 and POWER9 processors, this buffer is called  TMCAM and
consists of a content-addressable memory linked with the L2 cache, shared by eight hardware threads \cite{ibm_2018}.
Since the TMCAM is 8 KB in size, the available capacity for transactions running on the core(s) sharing a TMCAM is up to 64 cache lines. 

%\powerhtm detects conflicts at cache line granularity
%, which means that at most 64 memory addresses of shared objects can be accessed by some transaction. 

One of the main features of the POWER architecture is its extensive use of 
Simultaneous Multi-Threaded (SMT) which supports the execution of up to 8 threads per core (SMT-1,2,4,8).
When multiple threads run in SMT on the same core, they share the hardware resources available to that core. 
It is worth noting that SMT and \powerhtm are, in practice, conflicting features since the TMCAM is shared among co-located SMT threads. Therefore, a transactional program that takes advantage of SMT
will inherently reduce the available capacity to hardware transactions, which
degrades the effectiveness of \powerhtm. 
Most recently, POWER9 introduced an additional 512 KB read tracking structure, called L2 LVDIR and also shared among two cores, intended to support transactions with larger read sets \cite{ibm_2018}; however, the L2 LVDIR can only be used by up to two threads at any given time, which essentially makes it incompatible with workloads with large transactions that wish to use SMT.
Not surprisingly, we are not aware of any paper that has proposed a transactional system that exploits \powerhtm and achieves consistent speed-ups when running in SMT scenarios.

%joao: acima juntei mencao à nova estrutura para large read txs do P9. Esta claro e é boa ideia mencionar nesta forma critica aqui no segmento do SMT?

When running transactions in \powerhtm, the program can resort to an advanced suspend-resume mechanism. When a transaction is suspended, all its subsequent operations are executed non-transactionally, thus not tracked in the TMCAM. When the program eventually resumes the transaction any transactional conflicts that were detected during the suspend-resume interval take effect (and the transaction aborts). This is a useful mechanism to support programs that need to access control variables within a transaction's lifetime without aborting due
to conflicts when accessing those control variables.% thus enabling more advanced applications to be built on top of Power8's HTM.

Another advanced feature of \powerhtm is the support of a special kind of transactions called rollback-only transactions (ROT). In this mode, the TMCAM only tracks writes\footnote{Actually, due to implementation-specific reasons, the TMCAM can also track a small fraction of reads in a ROT \cite{ibm_P8_manual}.} while reads are performed as if they were not inside a transaction. This difference has two key consequences when we compare the behaviors of ROTs and regular HTM transactions. First, reads no longer contribute to spending the HTM capacity. Since reads are usually the most prevalent operation inside a transaction, the capacity of ROTs is improved massively relatively to regular HTM transactions.

\begin{figure}[t]
\centering
\includegraphics[width=0.5\textwidth]{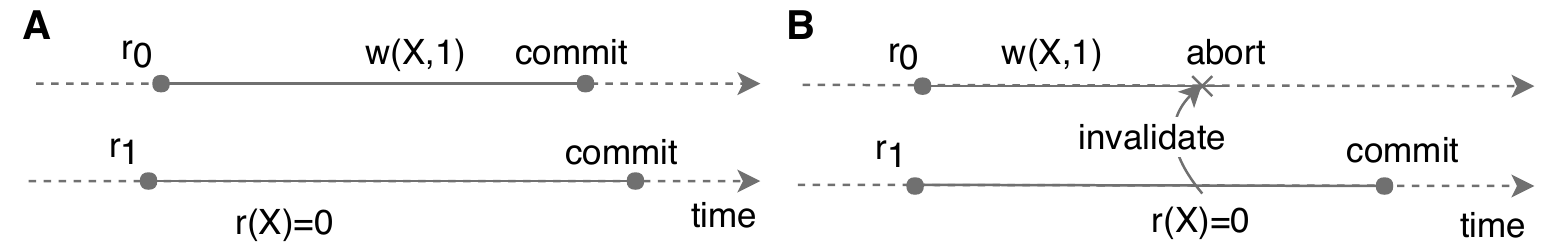}
\caption{Example A: write-after-read conflict is tolerated by ROTs. Example B: read-after-write conflict causes the writer ROT to abort.}
\label{fig:ROT-basic-conflicts}
\end{figure}

The second consequence is that, while concurrent ROTs can still abort due to write-write conflicts, some read-write conflicts are not guaranteed to be (and in general are not) detected, hence will not lead to aborts. Example A in Figure \ref{fig:ROT-basic-conflicts} shows an example of a write-after-read conflict between two ROTs that is tolerated. However, it should be noted that ROTs can still abort due to read-after-write conflicts. As illustrated in example B in Figure \ref{fig:ROT-basic-conflicts}, if a ROT $r_1$ writes to a location and a concurrent ROT $r_2$ later reads from the same location, $r_1$'s entry in the TMCAM will be invalidated and thus $r_1$ will abort.
Evidently, the weak semantics of ROTs do not guarantee serializability.
For that reason, the official documentation of \powerhtm clearly states that ROTs should only be used with code blocks that exclusively access thread-local data and may benefit from the ability to roll back their updates \cite{ibm_2018} -- hence the term ``rollback-only transactions''. 

As we show next, ROTs can actually be used with a different
purpose than intended and constitute a fundamental
building block to implementing a restricted form of hardware-supported SI
in \name.

%######################################################
\section{\name}
\label{sec:implementation}
%######################################################

The goal of \name is to build an SI implementation directly supported by
HTM.
The gains of \name design are two-fold: first, to benefit from the
fast transactional execution that an HTM delivers when
directly handling memory accesses and conflicts;
second, to take advantage of SI to avoid the tight capacity restrictions of
HTM.
Recent proposals have shown how to achieve this goal through modified hardware \cite{Litz:2014:SRT:2541940.2541952,Chen:2017:AGH:3079856.3080204}.
The key novelty of \name is that it relies on the commercially available IBM POWER8 and 9 architectures,
hence \name is ready to run on off-the-shelf hardware.

Accomplishing this design requires overcoming important challenges.
First, the available plain HTM transactions impose rigid strong semantics and
capacity limits. %Hence, to build a system with more relaxed restrictions, one needs to resort to weaker constructs that the hardware provides.
\name relies on ROTs as the main building block to execute transactions.
This inherently enables \name transactions to be capacity-bounded only by their write-sets, which
represents a decisive advantage with read-dominated and read-only transactions.
However, executing each transaction as a ROT that accesses shared memory is
unsafe, as it may yield serious anomalies that SI disallows.

To prevent the ROT-induced anomalies, the hardware ROT support needs to be
complemented with software-level instrumentation that enforces the hardware-supported execution to circumvent those anomalies.
The second challenge is, then, to ensure that such software instrumentation has
a reduced impact on the runtime performance of HTM.
Ideally, memory accesses should be handled directly by the HTM; 
any code instrumentation should only be allowed (and minimized) on the begin and commit stages -- especially for read-only transactions, which dominate many workloads.

Non-instrumented transactional accesses imply that
each transaction will directly access the cache-coherent memory, which is single-version
(i.e., two transactions that read from the same location simultaneously observe the same value).
This is incompatible with the original definition of SI, which relies
on a multi-versioned scheme to allow concurrent transactions to access
distinct isolated snapshots. An example is given in Figure
\ref{fig:SI-basics}, where, after $t_0$ has written to $X$ on its local snapshot, both $t_0$ and $t_1$ observe different values when they read from $X$.
The third and last challenge is, then, how to implement SI on a single-version
memory system. 

Since building a multi-versioned memory would require significant
software instrumentation on memory accesses, \name follows a different
approach: \name relies on the single-version memory system, which keeps transactional accesses
non-instrumented, and restricts the allowed executions to those that, under SI
rules, would not require keeping older data versions.
For instance, recalling the above example from Figure \ref{fig:SI-basics},
a correct execution (under SI) implies that $t_0$ and $t_1$ observe different values when accessing the same memory location. To deal with this, \name enforces that
one of the contending transactions (either $t_0$ or $t_1$) aborts.% -- thereby providing a \emph{restricted} implementation of SI.

%ricardo: paragrafo repetitivo?
%Summing up, \name relies on ROTs as its core speculative execution units, which perform direct 
%reads and writes to the (single-version) cache-coherent memory. This enables significant gains 
%in memory access latency when compared to software-based techniques; and
%dramatic gains in transactional capacity when compared to HTM.
%Since fully ensuring SI is not possible under such circumstances, \name
%implements a restricted form of SI, which allows a subset of the execution
%histories that are correct under SI.

The next sections describe \name in detail.
Section \ref{subsec:alg1} starts by discussing the semantics of encapsulating
transactions in ROTs, pointing out possible anomalies that are not accepted
under SI.
Section \ref{subsec:alg2} then complements ROTs with the necessary software instrumentation to ensure that allowed executions are correct under SI.
Section \ref{subsec:alg3} then describes how read-only transactions may be optimized.

%joao: check this
We describe \name as a support for general-purpose transactional memory programs that, within each transaction may read and write to pre-allocated memory locations, indexed by their virtual address.
Among other uses, \name can be integrated as a concurrency control mechanism in IMDBs, including IMDBs that stores named records that are accessed by a set-oriented language (like SQL), making use of efficient indexes \cite{Tu:2013:STM:2517349.2522713}. 
%; however, we leave that extension
%out of the scope of the paper.

\subsection{ROTs as the building block of \name}
\label{subsec:alg1}

%Although the original purpose of ROTs is to support speculative execution
%of code sections that do not access shared memory, \name violates that practice
%and encapsulates each transaction on a ROT.
\name encapsulates each transaction in a ROT by preceding each transaction's code by a \rotbegin instruction  and committing the transaction with \rotcommit.
Yet, as Section \ref{sec:background} discussed, the semantics of ROTs are
unsuitable to transactional programs, as they may yield serious correctness
anomalies.
Still, it is also true that ROTs implicitly ensure some key consistency
properties that are shared with SI.
Namely:

\begin{itemize}
\itemsep 0em 
\item Since \powerhtm keeps track of each ROT's write set, the underlying
  hardware 2PL implementation detects write-write conflicts and resolves
  them by aborting (at least) one of the contending ROTs.
  This implicitly satisfies SI's restriction that, when two concurrent transactions have overlapping write sets, one of them should not be allowed to commit.%\footnote{To be precise, SI defines that, when write-write conflicts arise, either the first writer or the first committer wins \cite{**}. In contrast, \powerhtm aborts the transaction that wrote to the contended location. However, as Section \ref{***} shows, this difference has no relevant impact on the correctness of \name.}

%\item Since \powerhtm does not keep track of the reads of ROTs,
%  the capacity limits imposed by \powerhtm are now reduced to the write-set
%  of each transaction, which is the minimal footprint that one expects
%  on an SI implementation. This represents a substantial improvement on
%  the available capacity, since read sets no longer contribute to filling up
%  the TMCAM.

\item  Executions where a ROT writes
  to a location that has previously been read (and not written) by an ongoing
  concurrent ROT are not treated as conflicts. While this was not allowed
  by the serializable 2PL implementation of plain HTM, it is allowed under SI.
  Of course, as Section \ref{sec:background} discusses, ROTs still treat some
  read-write situations as conflicts. This reflects the fact that \name is
  a single-version implementation of SI.

%  If should be noted that, in theory, SI avoids any other read-write conflicts, whereas \powerhtm treats read-after-write cases as conflicts, and may even
%  decide to abort in arbitrary situations due to its best-effort nature (as Section \ref{sec:background} details. Hence, with respect to read-write conflicts,
%  ROTs are a restricted approximation of SI.

\end{itemize}

\begin{figure}[t]
\centering
\includegraphics[width=0.3\textwidth]{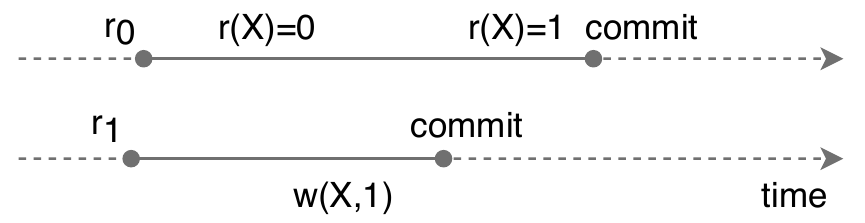}
\caption{Example of two concurrent ROTs contending on shared data with an anomaly that is not allowed under SI}
\label{fig:ROT-tx-anomaly}
\end{figure}

Therefore, by running transactions as ROTs, \name gets the above SI properties
for free from the hardware.
However, using ROTs to encapsulate transactions that concurrently access shared data may yield \emph{dirty read} anomalies \cite{ANSI:1992:ANSd}, which are not allowed under SI \cite{Berenson:1995:CAS:223784.223785}.

To illustrate these anomalies consider the example in Figure \ref{fig:ROT-tx-anomaly}. In this example, two concurrent ROT-encapsulated transactions, $r_1$ and $r_2$ access shared variable $X$. Since $r_2$ writes to $X$ after $r_1$ reads $X$ (a write-after-read case), no conflict is detected, hence both ROTs are allowed to continue running. However, since $r_2$'s write is performed in the actual shared location, this write is not isolated in $r_2$'s conceptual snapshot, as SI mandates.
Instead, the write is visible to $r_2$ when $r_2$ later reads from $X$.
Recall that, in an execution that is correct under SI, the second read by $r_2$ should return the value of $X$ that was committed when the (isolated) snapshot was initially created -- clearly, the above execution with ROTs results in a dirty read, which violates the requirement of isolated snapshots in SI.

%joao: is there any other distinct anomaly to depict? (that is not the typical SI write-skew)?

The next section shows how \name prevents dirty reads on ROT-encapsulated transactions.

%######################################################
\subsection{Base algorithm}
%######################################################
\label{subsec:alg2}

%joao: falta rever para incluir timestamps na explicacao

\begin{figure}[t]
\centering
\includegraphics[width=0.4\textwidth]{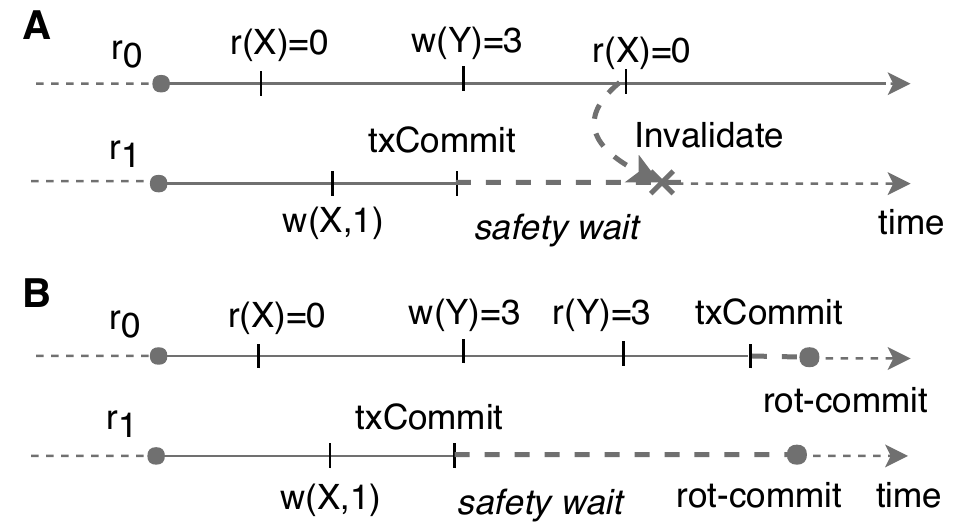}
\caption{Examples of the safety wait as a means to prevent dirty read anomalies. Example A: a dirty read is effectively
  prevented by having $r_1$ wait until $r_0$ performs the problematic read. Example B: after a safety wait,
  $r_1$ commits without causing dirty reads.}
\label{fig:quiescent-commit}
\end{figure}

Recall the example in Figure \ref{fig:ROT-tx-anomaly} illustrating the dirty read anomaly induced by encapsulating transactions in ROTs.
The key insight behind \name is that, if ROT $r_1$ had waited for a \emph{sufficiently long} time before issuing the \rotcommit instruction, the anomaly would not
have occurred.

Suppose that $r_1$ had waited until $r_0$ had issued its last read, $r(X)$,
as Figure \ref{fig:quiescent-commit} illustrates in example A.
Since this read targets a location that is currently in $r_1$'s write-set, $r_0$'s read invalidates 
$r_1$'s write entry in the TMCAM and $r_1$ aborts.
Consequently, $r_1$'s write is rolled back, thus $r_0$ reads the original (and correct) value ($X=0$).
Therefore, this waiting prevents the dirty read anomaly by aborting the writer transaction.

It should be noted that $r_1$ can only be sure that it can safely commit without
incurring dirty read anomalies on $r_0$ after $r_0$ has performed its last access (and $r_1$ has survived
each read access by $r_0$). Example B in Figure \ref{fig:quiescent-commit} illustrates an
alternative execution where $r_0$ does not read from locations written by $r_1$ and, thus,
$r_1$ can safely commit after waiting $r_0$'s last read.

We should remark that, in both examples in Figure \ref{fig:quiescent-commit}, there is
a cost to pay for correctness: in example A, the writer transaction aborts; 
in example B, the writer transaction spends significant time spinning. 
As we describe next, \name trades these costs for important improvements in capacity. As Section \ref{sec:evaluation} evaluates, the gains clearly outweigh these costs.

%\name generalizes the above insight to conservatively cover any scenario
%where an arbitrary number of ROTs are running concurrently.

More generically, when a ROT-encapsulated transaction, $r$ that wrote to a given 
location, $X$, completes (i.e. performs its
last memory access before entering the commit phase), $r$ should wait until
any active transaction, $t$, that has read $X$ before $r$'s update is guaranteed to
have performed its last read of $X$.
If $r$ only issues \rotcommit after that condition is guaranteed, then none of $r$'s
writes will induce dirty reads on any $t$.

However, precisely determining the earliest instant when such a guarantee
is met is impractical.
Firstly, it is usually not possible to predict the future reads of a transaction.
Second, determining which ROTs have previously read $X$ would imply keeping
track (at the software-level) of each read, requiring
prohibitive read instrumentation.
Hence, \name adopts a conservative approach: $r$ assumes that
every \emph{active} ROT may have read from at least one location of $r$'s
write set; moreover, $r$ waits until each such concurrent transaction has
completed, thus it will not issue any more memory requests.
When this condition is satisfied, we say that $r$ is \emph{safe to commit}.

%joao: check this (changed names do HTM instructions)
\begin{algorithm}[t]
  \caption{Transaction begin and end}
  \label{alg:simple}
  \scriptsize
\begin{algorithmic}[1]
  \State int $state[N] = \{inactive, .., inactive\}$
  %\LineComment{each entry can be inactive=0, completed=1, active>1}
\\
  \Function{TxBegin()}{}
    	\State $state[tid] \gets currentTime()$\label{alg1:setActive}
        \State $sync()$ 
    	%\LineComment{show our transaction as active}\\
        \If{$HTMBeginROT$}
        \State \Return
        \EndIf{}
        %\LineComment{successfully started a tx}
  \EndFunction
  \\
  \Function{TxEnd()}{}
  	\State $HTMSuspend$
        %\LineComment{Suspend the ROT}
        \State $state[tid] \gets completed$\label{alg1:setCompleted}
        \State $sync()$
        \State $HTMResume$
        %\LineComment{signal the tx is committing}
        \State $snapshot[0..N-1] \gets state[0..N-1]$
        \label{alg1:statesnapshot}
        %\LineComment{Take a snapshot of all txs}
        \For{$c : 0..N - 1, c \neq tid$}\label{alg1:wait1}
          \If{$snapshot[c] > 1$}
        %\LineComment{quiesce all current active ROTs}
    	    \State \textbf{wait while} $snapshot[c] = state[c]$\label{alg1:wait2}
          \EndIf
        \EndFor
        
        \State $HTMEnd$
        \label{alg1:tmend}
        %\LineComment{resume ROT and end transaction}
        \State $state[tid] \gets inactive$\label{alg1:inactive}
        %\LineComment{set tx state to inactive}
  \EndFunction
\end{algorithmic}
\end{algorithm}

Algorithm \ref{alg:simple} shows the software-level instrumentation
that \name adds to implement the safety wait.
\name maintains a shared array where each thread publishes its current
state, which can either be not running any transaction (\texttt{inactive=0}), running a transaction (any value greater than 1), or completed and waiting for a safe commit (\texttt{completed=1}).

Before starting the ROT on \powerhtm, a transaction
announces that it will become active (line \ref{alg1:setActive})
by setting its state to the current system timestamp (in clock cycles).
Conversely, when a transaction has completed and wishes to commit, it
sets its state to \texttt{completed} (line \ref{alg1:setCompleted}) and then waits until every
other active transaction, $t$, leaves that state. After $t$ leaves
that state, the corresponding thread's state will change to \texttt{inactive}, eventually starting another transaction and becoming \texttt{active} again with a
higher timestamp, and later switching to \texttt{completed}, and so forth.

The waiting condition in lines \ref{alg1:wait1}-\ref{alg1:wait2} spins until one of such options is observed for each other thread. Once that happens, the waiting transaction can finally commit its ROT in \powerhtm and announce itself as \texttt{inactive} (line \ref{alg1:inactive}).

Whenever a transaction's state is set, after the beginning of a new transaction (line \ref{alg1:setActive}) and after suspending (line \ref{alg1:setCompleted}), we need to ensure that the change propagates to all concurrent transactions. To do so, we enforce a full memory barrier (\emph{sync}) after the state change.

One relevant implementation detail is that all updates to the thread's
entry on the shared state array are performed in non-transactional mode.
%More concretely, the initial update of the transaction's state is performed
%before the corresponding ROT begins. 
If these updates happened inside an active ROT, the transactional buffer would be occupied with one unnecessary write and, most importantly, the ROT would abort whenever other transactions read this transaction's state in the shared array.
%For the same reason, before signaling that the transaction is \texttt{completed}, the ROT is suspended so that the state update ocurs in non-transactional mode.

%joao: alguma razao para o resume so acontecer depois da quiescence terminar? ou podia acontecer antes e assim a espera acontecer em modo transacional? (A espera em modo transacional tem a vantagem da tx ser abortada mais rapidamente em caso de read-after-write, e sempre é mais intuitivo para o leitor entender.)

%If is easy to show that, given a set of concurrent transactions,
%every such transaction will become safe to commit at the same instant.
%Of course, the exact timings by which each transaction determines that
%it is safe to commit may diverge across transactions.

%We use a quiescence barrier similar to RCU's \cite{rcu} in the suspend-resume phase in order to achieve a memory consistency model that is much stronger than what ROTs alone provide. Before ending a ROT's execution, we suspend the ROT, we force it to wait for any other concurrently running ROTs to arrive at the commit step, we resume the ROT and commit it. This ensures that some ROT will always have the chance to be aborted by conflicts triggered from concurrent ROTs reads or writes, as shown in Figure \ref{fig:si-htm-execution}.

%\input{code}

%######################################################
\subsection{Read-only fast path and the fall-back path}
%######################################################
\label{subsec:alg3}

\name has alternative paths besides the algorithm described so far,
which Algorithm \ref{alg:complete} presents.

\algnewcommand{\algorithmicgoto}{\textbf{go to}}
\algnewcommand{\Goto}[1]{\algorithmicgoto~\ref{#1}}
\begin{algorithm}[t!]
  \caption{Extension with SGL and RO paths}
  \label{alg:complete}
\scriptsize
\begin{algorithmic}[1]
  \Function{SyncWithGL()}{}
    \State $state[tid] \gets currentTime()$ \label{alg2:set-state}
    \State $sync()$
    \If{$globalLock.isLocked()$}
      \State $state[tid] \gets inactive$
      \State \textbf{wait while} $globalLock.isLocked()$
      \State \Goto{alg2:set-state}
    \EndIf
  \EndFunction
  \\
  \Function{TxBeginExt(boolean isRO)}{}
    \If{isRO}
        \State $SyncWithGL()$
        \State \Return
      %\LineComment{wait until the GL is unlocked}
    \Else
      \While{$(retries$ - - $ > 0)$} \label{alg2:retries}
        %\LineComment{times to try starting a new tx}
        \State $SyncWithGL()$
        \If{$HTMBeginROT$}
        \State \Return
        \EndIf{}
      \EndWhile
    \State $state[tid] \gets inactive$
%    \LineComment{Fallback to GL if retries end}
    \State $globalLock.lock(tid)$\label{alg2:sgl}
    \For{$c : 0..N-1, c \neq tid $}
    	%\LineComment{wait for all concurrent txs to complete}
    	\State \textbf{wait while} {$state[c] \neq inactive$}
	\EndFor
    \EndIf
  \EndFunction
  \\
  \Function{TxEndExt()}{}
  	\If{$globalLock.isLocked(tid)$}
    	\State $globalLock.unlock()$
%        \LineComment{Release the GL if tid has it}
    \Else
        \If{isRO}\label{alg2:ro-commit}
        	\State $lwsync()$ \label{alg2:lwsync}
        	\State $state[tid] \gets inactive$
        \Else
    		\State $TxEnd()$
        \EndIf
    \EndIf
  \EndFunction
\end{algorithmic}
\end{algorithm}

One important path is the fast path for read-only transactions.
In the context of \name, we define a read-only transaction as one that
performs no writes on shared data locations.
Note that a read-only transaction is allowed to update thread-private
memory locations (such as its local stack).
When a transaction is launched in \name, an argument specifies whether
the transaction is read-only or not.
We assume this parameter is set by the programmer or by some automatic
tool (e.g. a compiler).

Since read-only transactions perform no shared updates, they are not prone
to cause dirty reads. Therefore, they may safely skip the safety waiting and
immediately commit upon completion (line \ref{alg2:ro-commit}).
Of course, read-only transactions still need to announce their state changes,
so that other read-write transactions
can know how to coordinate their safety waitings.
%Again, after a transaction's state is set at the beginning of either fallback path we need to make sure concurrent transactions see this change by enforcing a full memory barrier with \textit{sync}. 
Finally, when a read-only transaction ends we must ensure that all shared memory reads were performed before the state is set. This is accomplished using a light-weight barrier \textit{lwsync} issued at Line \ref{alg2:lwsync}.

Another alternative path is the fall-back path when a read-write transaction is not able to commit after a number of retries.
This can happen for a number of reasons, ranging from transactions
whose write-sets exceeds the available capacity, frequent
aborts under high contention, to transactions
issuing instructions that are illegal on \powerhtm, among others.
To ensure progress under such situations, transactions in \name resort
to a traditional single global lock (SGL) fall-back path after having aborted
too many consecutive times (line \ref{alg2:retries}).
We should note that, upon acquiring the SGL, the transaction cannot proceed immediately
since other concurrent transactions may still be actively running. Hence, the SGL holder
first waits until no other active transaction exists (lines \ref{alg2:sgl}).
While the SGL is locked, no other
transaction is allowed to proceed\footnote{It should be noted that the early subscription scheme
that is usually used with regular HTM (which precludes the initial wait after the SGL lock) is not possible in \name since read-only transactions run
non-transactionally and ROTs do not detect write-after-read conflicts.}.

%######################################################
\subsection{Correctness}
%######################################################
\label{sec:correctness}

%joao: check this
%This section presents a sketch of proof that,
%\name implements a restricted form of SI.
%In other words, 
%Even though
%\name precludes some execution
%histories that SI enables, 
In this section, we show that any execution history that \name allows
is correct under the original definition of SI 
\cite{Berenson:1995:CAS:223784.223785}.
%This property is
%fundamental to the applicability of \name.
This implies that any existing application that is serializable under SI
may be directly executed under \name and will retain its correctness.
Moreover, it means that the techniques and tools that have been proposed
to analyze and fix programs to run under SI without anomalies (e.g., \cite{Fekete:2005:MSI:1071610.1071615,Litz:2014:SRT:2541940.2541952}) are also
applicable to \name.

We note, however, that \name assumes that any access to shared memory locations 
is included within a \name transaction (i.e., between \texttt{TxBeginExt} and \texttt{TxEndExt}).
This is similar to the weak atomicity model \cite{Martin:2006:STM:1251552.1251589}, albeit employed in the context of SI. 
We also remark that 
\name prevents inconsistent reads, in the sense that any transaction in \name
(even one that eventually aborts) must see (all and only) the effects of transactions that committed before they started.

However, we highlight that \name also prevents inconsistent reads from 
transactions that eventually abort, which is a stronger guarantee that
SI does not require.
It is straightforward to prove this property by recalling the semantics 
of \powerhtm. Let us suppose, by contradiction, that some transaction, $t$, performs an inconsistent read, i.e. reads a value written by some transaction that has or will abort. This would mean that $t$ was able to read a value written by another transaction that
had not yet been committed, which contradicts \powerhtm's semantics.
A consequence of this property is that, 
for those applications \emph{that are serializable under SI} (i.e., are guaranteed to run under SI without incurring SI-related anomalies)
\name will not yield the undesirable side-effects that may arise in TM implementations
that allow inconsistent reads \cite{Guerraoui:2008:Opacity}.
Therefore, \name can safely run such applications
in non-sandboxed environments.

To show that any execution history that \name allows is correct under SI,
the following sketch of proof addresses each restriction from the operational definition of SI 
\cite{Berenson:1995:CAS:223784.223785},
explaining why \name satisfies all of them.

\vskip 5pt

\textbf{R1: Each transaction reads data from a snapshot
of the (committed) data as of the time the transaction started, called 
its \emph{Start-Timestamp}.}
  
\begin{proof}

Rephrasing the above restriction, any transaction $t$ reads from a snapshot that reflects the writes by the most recently committed transactions whose \emph{Commit-Timestamp} precedes $t$'s Start-Timestamp.
The above restriction is guaranteed 
if, for any transaction $t$ that successfully commits (in \name), we define its Commit-Timestamp
as the instant where $t$ completes taking a snapshot of each thread's state (line \ref{alg1:statesnapshot} in Alg. \ref{alg:simple}); 
i.e., just before performing its safety wait.

Consider a pair of transactions, $t_w$ and $t_r$, where: transaction $t_w$ writes to a given location, $o$, and $t_w$ eventually commits;
$t_r$ starts after $t_w$ commit and reads from $o$; $t_w$ is the last transaction to write to $o$ before the Start-Timestamp of $t_r$.
Assume, by contradiction, that $t_r$ observes a version, $v*$, different from the one produced by $t_w$. 
This would be possible only in the following three cases:

Case a:  $v*$ is produced by a not yet committed transaction. This is impossible, since it implies that $t_r$ read uncommitted values, which is prevented by the  \powerhtm. Note that this would be a violation of restriction \emph{R4}, which we address later.

Case b: $v*$ is produced by a transaction, $t'$, that committed after $t_r$ started; i.e., the Start-Timestamp of $t_r$ precedes the Commit-Timestamp of $t'$.
Clearly, $t_r$ can only read $v*$ after $t'$ has issued \rotcommit.
%If $t_r$ reads $v*$ before $t_{w}$ issues \rotcommit, then a read-after-write conflict
%arises (between the read by $t_r$ and the write by $t'$), which \powerhtm solves by aborting
%$t_r$ or $t'$ (in case $t_r$ is also a ROT), which contradicts the hypothesis that both transactions commit.
%If, alternatively, $t_r$'s read occurred after $t'$ issues \rotcommit, 
This implies
that $t'$ completed its safety wait before $t_r$ read $o$. However, by hypothesis, $t_r$'s Start-Timestamp is earlier than
the Commit-Timestamp of $t'$, consequently $t'$ observed that $t_r$'s state was active before 
$t'$ initiated its safety wait. Therefore, $t'$ could only conclude its safety wait after 
$t_r$ committed, which contradicts the hypothesis that $t_r$ read after $t'$ executed \rotcommit.

Case c: $v*$ is produced by a committed transaction, $t'$, whose Commit-Timestamp is earlier than $t_w$'s Commit-\\Timestamp.
By restriction \emph{R5} (described later), the only case where both $t'$ and $t_w$ are able to commit is if the Commit-Timestamp of $t'$ precedes the Start-Timestamp of $t_w$.
Since, by hypothesis, both transactions commit, then it is easy to prove that $t_w$'s write to $o$ occurred after $t'$ executed \rotcommit (otherwise, a write-write conflict would arise and \powerhtm would abort one writer) and, consequently, after the write by $t'$. Since, by hypothesis, $t_r$ starts after the Commit-Timestamp of $t_w$, then $t_r$ will necessarily observe the most recent write ($t_w$'s write), which contradicts case c's hypothesis.
% As a first hypothesis, let us consider that the Commit-Timestamp of $t_w$
% precedes $t$'s Start-Timestamp; however, $t$ gets $v_0$ when reading from $o$, thus violating the above restriction.
% Clearly, $t$'s read occurred before $t_{w}$'s write; otherwise, the semantics of \powerhtm dictate that $t$ would have read $v_1$ (since, by hypothesis, $t_w$ did not abort). 
% However, as defined above, $t_{w}$'s Commit-Timestamp denotes a time in which $t_w$ has already executed all its writes. This implies that $t$'s read preceded
% $t$'s Start-Timestamp, which is a contradiction.
\end{proof}

\begin{figure}[t]
\centering
\includegraphics[width=0.3\textwidth]{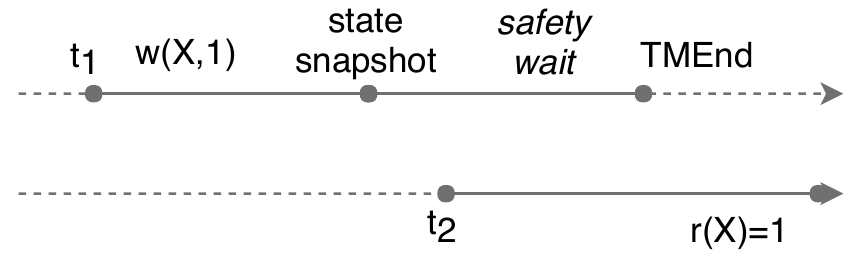}
\caption{Example illustrating why selecting the Commit-Timestamp of $t_1$ is defined as 
the time where $t_1$ completes reading each thread's state, rather than the time at which \rotcommit occurs.}
\label{fig:commit-instant-proof}
\end{figure}

As a side note, we
explain the rationale behind defining the Commit-Timestamp
as defined above, instead of the instant in which \rotcommit is performed (line \ref{alg1:tmend} of Alg. \ref{alg:simple}).
To illustrate why the alternative definition is not appropriate, consider the example in Figure \ref{fig:commit-instant-proof}. Transaction $t_2$ reads the value that $t_1$ wrote 
(and committed) to $x$. However, $t_2$ began \emph{before} $t_1$ 
executed \rotcommit. Hence, considering the moment at which a 
transaction performs \rotcommit as its Commit-Timestamp
would contradict the previous SI restriction.

\textbf{R2: A transaction running in SI is never
blocked attempting a read as long as the snapshot data from
its Start-Timestamp can be maintained.}

\begin{proof}Trivially ensured by \powerhtm's semantics, which never block upon memory accesses.\end{proof}

\textbf{R3: A transaction's writes are also reflected in its local snapshot.} 
\begin{proof}
Since update transactions are executed in ROTs, the semantics of \powerhtm trivially ensure this restriction. 
%A write performed by
%a ROT is cached in the local transactional buffer, and any reads by the
%corresponding thread will observe the buffered value.
\end{proof}

\textbf{R4: Updates by other transactions active after the transaction Start-Timestamp are invisible to the transaction.} 
%As Section \ref{sec:implementation} discusses intuitively,
%the safety wait before each rot-commit of writer transactions ensures
%that this property is always met.
\begin{proof}
Let us assume that a transaction $t_w$ writes on a given
location and another transaction, $t_r$, reads the same location.
Further, let us assume that $t_w$ is active (i.e., $t_w$ has not reached its Commit-Timestamp, as defined in R1) after $t_r$ started.
If $t_w$ still has not executed \rotcommit when the read occurs, then \powerhtm
invalidates $t_w$'s write (thus aborting $t_w$) and, thus, $t_r$ reads the previous value, which satisfies R3.

Alternatively, let us suppose, by contradiction, that $t_w$ had already committed when $t_r$ reads, then $t_r$ would see $t_w$'s update. However, since $t_w$ had already performed \rotcommit,
then we know that $t_w$ had previously completed its safety wait. This
implies that every transaction that had started before $t_w$'s Commit-Timestamp has already
completed; since $t_r$ subsequently performs a read, then $t_r$ has started after $t_W$ has committed, which contradicts the initial hypothesis.
\end{proof}

\textbf{R5: A transaction $t_1$ can only commit if no other transaction, $t_2$, with a Commit-Timestamp in $t_1$'s execution interval [Start-Timestamp, Commit-Timestamp] wrote data that $t_1$ also wrote.}
\begin{proof}
Suppose, by contradiction, that transactions $t_1$ and $t_2$ write to a common location, $o$. Further, 
suppose, without loss of generality, that $t_2$ commits before $t_1$ does.
%suppose that $t_2$ has committed while $t_1$ is about to commit. 
Since, by hypothesis, $t_1$'s Start-Timestamp precedes $t_2$'s Commit-Timestamp, $t_2$ must have observed that the state of $t_1$ was not \texttt{inactive} before $t_2$ entered its safety wait. 
Therefore, $t_2$ had to wait until $t_1$ issued all its memory accesses and completed (since, by hypothesis, $t_1$ did not abort). 
This implies that $t_1$ wrote to $o$ before the write to the same location by $t_2$ was committed in hardware, which is a write-write conflict that \powerhtm solved by aborting either $t_1$ or $t_2$, which 
contradicts the hypothesis that both $t_1$ and $t_2$ commit.
\end{proof}
%Let's consider, by contradiction, that two concurrent writer %transactions, $t_0$ and $t_1$ write to a common location and both successfully commit.
%The safety wait guarantees that both $r_0$ and $r_1$ can only commit
%after both have completed, thus after both issued the conflicting write.
%Since \powerhtm aborts at least one writer when two concurrent ROTs write to the sama location either $t_0$ or $t_1$ has aborted, which contradicts the initial hypothesis. 
%\end{proof}

%Regarding the last restriction, we remark that \name adopts a \emph{last
%  writer wins} rule, instead of the \emph{first committer (or writer) wins}
%rule originally formulated in SI \cite{Berenson:1995:CAS:223784.223785}.
%This implies that, in some executions, \name will produce
%different execution histories than an exact implementation of SI.
%Yet, it does
%not change the fundamental result that this section proves: any execution history that \name allows is also allowed under SI.

%This condition provides important guarantees regarding whether other
%ROTs may observe or not the updates that $r$ has performed.
%More precisely, once $r$ is safe to commit (and has not aborted meanwhile),
%it is guaranteed that: i) any ROT whose snapshot is earlier than the instant
%at which $r$ rot-commits will not observe any update issued by $r$;
%ii) any ROT whose snapshot is created after 

To conclude, we complement the above sketch of proof with correctness arguments for the SGL fall-back path scenario.
As Section \ref{subsec:alg3} briefly explains, after a transaction acquires the SGL, it waits until ongoing transactions finish (lines 14-15, Alg.2) before it starts executing.
Conversely, any other transaction that may try to start will observe
that the SGL is taken and wait (lines 4 and 7, Alg.2) until that condition changes. 
Therefore, when the thread holding SGL runs its transaction, no other threads have active transactions and any previous transaction that had issued writes must have already committed or aborted.
Consequently, it is easy to show that the previous restrictions apply in the scenario where one transaction runs in the SGL fall-back path.
\section{Evaluation}
\label{sec:evaluation}

\setlength{\abovecaptionskip}{5pt}
\setlength{\belowcaptionskip}{5pt}
\begin{figure*}[t]
%\centering
%\includegraphics[width=0.25\textwidth]{plots/hashmap-hirocap-lowcon-u10-throughput}
%\includegraphics[width=0.245\textwidth]{plots/hashmap-hirocap-lowcon-u10-aborts}
\includegraphics[width=0.48\textwidth]{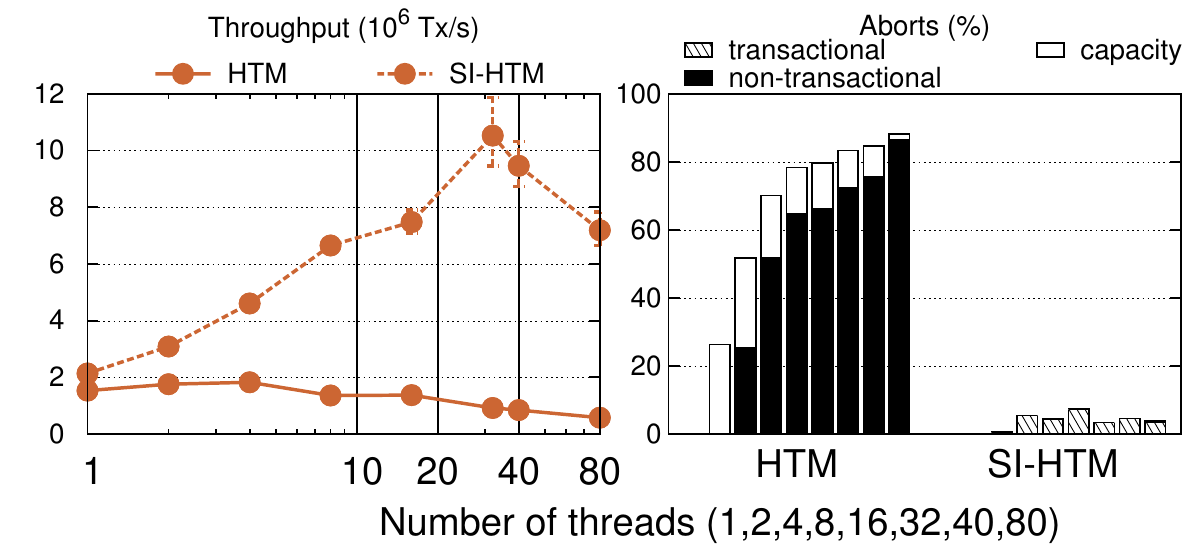}
%\caption{Hashmap 90\% large read-only txs, low contention}
\includegraphics[width=0.48\textwidth]{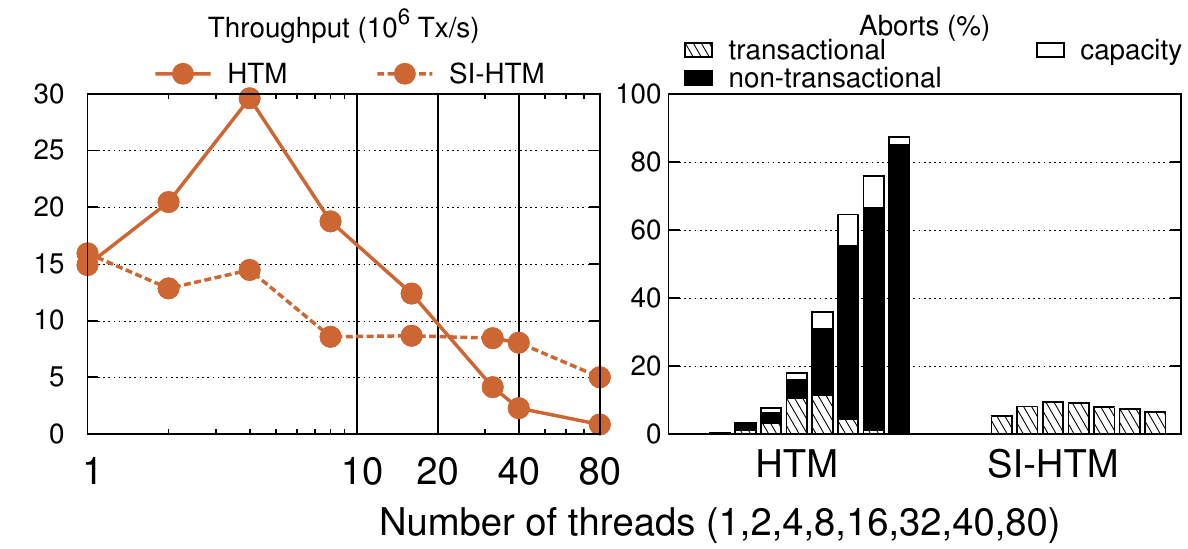}
\caption{Hash-map 90\% large read-only txs, low (left) and high (right) contention}
\label{plot:hashmap:largeRO:low}

%\includegraphics[width=0.25\textwidth]{plots/hashmap-hirocap-lowcon-u50-throughput}
%\includegraphics[width=0.245\textwidth]{plots/hashmap-hirocap-lowcon-u50-aborts}
%\caption{Hashmap 50\% large read-only txs, low contention}
%\includegraphics[width=0.25\textwidth]{plots/hashmap-hirocap-hicon-u50-throughput}
%\includegraphics[width=0.24\textwidth]{plots/hashmap-hirocap-hicon-u50-aborts}
\includegraphics[width=0.49\textwidth]{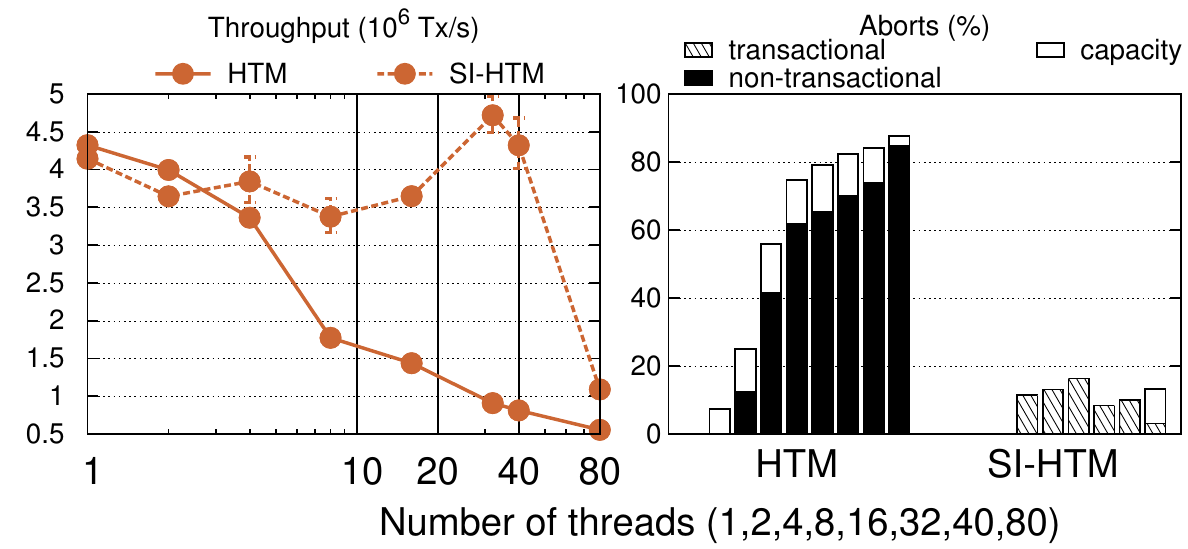}
%\caption{Hashmap 90\% large read-only txs, low contention}
\includegraphics[width=0.49\textwidth]{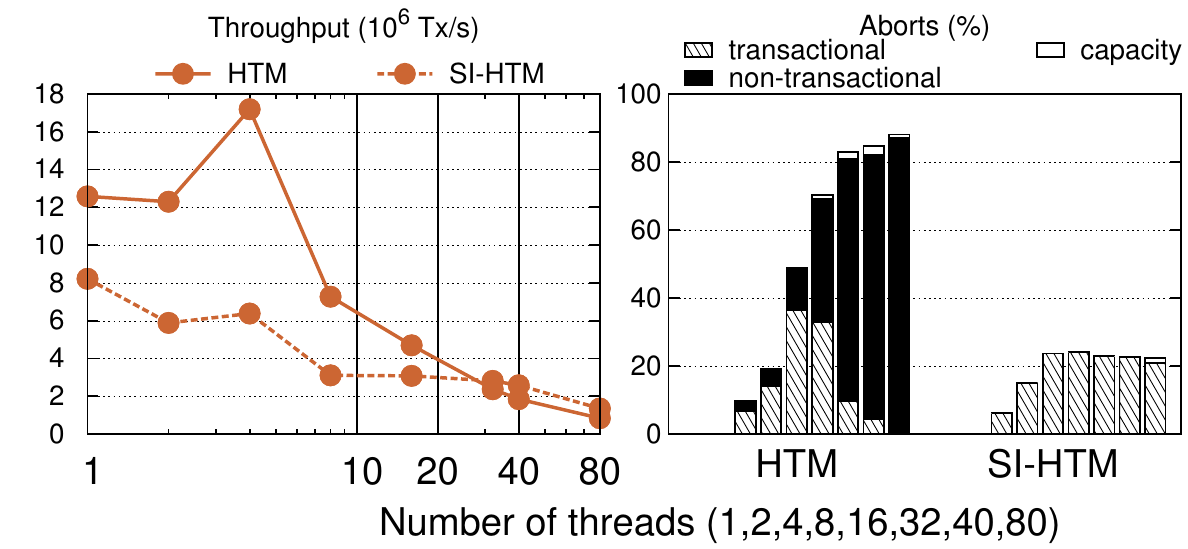}
\caption{Hash-map 50\% large read-only txs, low (left) and high (right) contention}

\includegraphics[width=0.48\textwidth]{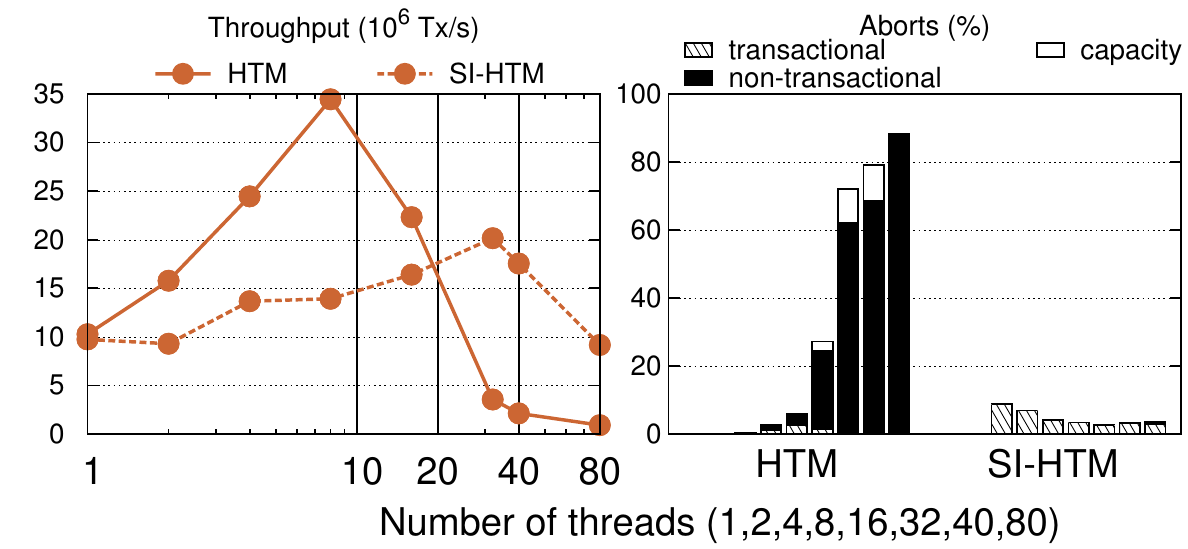}
%\caption{Hashmap 90\% large read-only txs, low contention}
\includegraphics[width=0.48\textwidth]{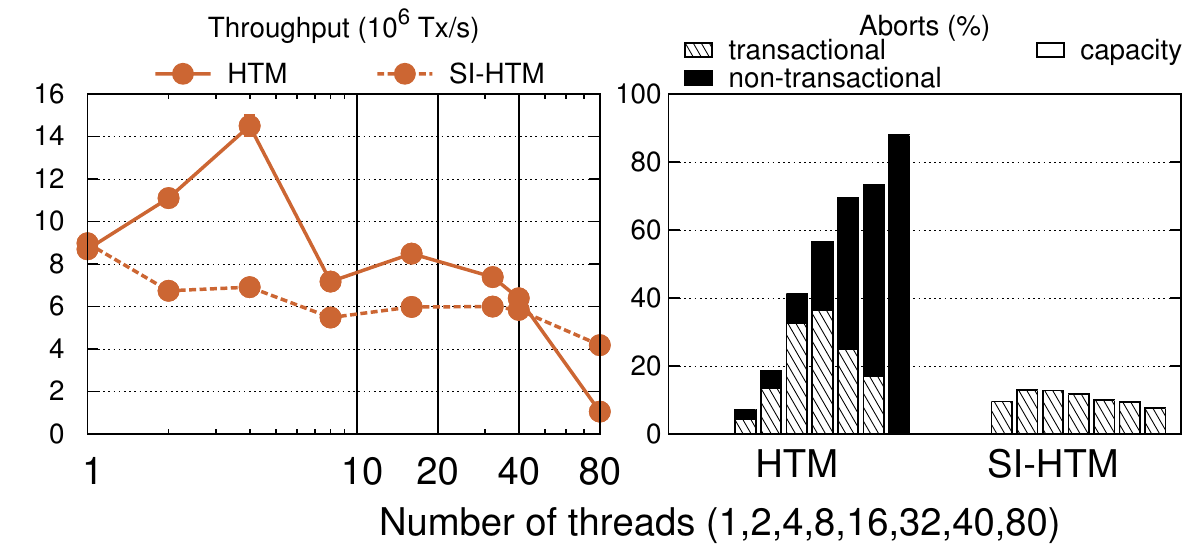}
%\includegraphics[width=0.25\textwidth]{plots/hashmap-lowcap-lowcon-u10-throughput}
%\includegraphics[width=0.24\textwidth]{plots/hashmap-lowcap-lowcon-u10-aborts}
%\caption{Hashmap 90\% small txs, low contention}
%\includegraphics[width=0.25\textwidth]{plots/hashmap-lowcap-hicon-u10-throughput}
%\includegraphics[width=0.245\textwidth]{plots/hashmap-lowcap-hicon-u10-aborts}
\caption{Hash-map 90\% small txs, low (left) and high (right) contention}
\label{plot:hashmap:small:high}
\end{figure*}

%Recalling Section \ref{sec:implementation}, 
\name has distinct features that have the potential to contribute to effective performance improvements.
More precisely, when compared with plain HTM, \name potentially offers the following benefits: i) update transactions with much larger memory footprints may run and commit without exhausting the HTM capacity; ii) read-only transactions run non-transactionally, hence exhibit lower begin/commit overheads and have unlimited capacity; iii) as a corollary of the previous outcomes, it becomes feasible to co-locate more parallel transactions on  a common SMT core; iv) since \name provides weaker correctness 
guarantees than plain HTM, \name allows higher concurrency.

The main goal of this evaluation is to understand, for a wide range of scenarios and workloads, the performance and scalability gains of \name, when compared to relevant HTM- and software-based concurrency control mechanisms.
in this study we aim to evaluate the effective benefits that each factor above (i to iv) contributes to the global outcome of \name, as well as the real performance costs of the quiescence phase component of \name.

In order to answer these questions, we deploy \name on a IBM Power8 system with one 8284-22A processor of 10 cores with SMT-8 (i.e., up to 8 hardware threads per core). We use a hash-map micro-benchmark to compare the behavior of \name with the pure HTM baseline; and TPC-C \cite{TPCC} as a real-world application benchmark, which we use to compare \name with a relevant set of state-of-the-art concurrency control systems.

%deves enumerar quais os sistemas com os quais te comparas. explicar que, na seccao da micro-benchmark, focaremos apenas a comparacao com HTM pois é a principal baseline e assim é fácil compreender os resultados.

\subsection{Hash-map benchmark}

The hash-map benchmark consists of a simple transactional hash-map implementation, where clients can perform lookup, insert and remove operations. A read-only transaction performs a lookup operation and a read-write transaction performs an insert, or a remove operation if the last transaction on that thread was an insert.
This synthetic benchmark allows us to study different workload scenarios that cover distinct combinations between the orthogonal dimensions of transaction footprint and contention.

Regarding the transaction footprint dimension, the
number of elements that initially populate the hash-map can be in one of two modes:
a \emph{large} transaction footprint mode, where the hash-map size is such that each bucket has, on average, a list of 200 elements (hence, operations on a key in that bucket may need to read
from 200 cache lines at most to find the target element, which easily leads transactions to exceed \powerhtm's capacity); and a 
\emph{short} transaction footprint mode, where each bucket has, on average, 50 elements (thus most operations find the target element without exceeding \powerhtm's capacity).
Since the available capacity, both in HTM and in \name, depends on
the read/write ratio, we further distinguish the (large vs. short) transaction footprint dimension with the read/write ratio. Thus, in total we consider 3 scenarios: a large-footprint scenario dominated by read-only transactions (90\% read-only transactions vs. 10\% update transactions); a large-footprint scenario 50\% read-only vs. 50\% update transactions; and a short-footprint scenario 
that mixes 90\% read-only transactions and 10\% update transactions. (We omit the short/50\%:50\% case for space limitations, as it adds no relevant findings.)

Concerning the orthogonal dimension of contention, it can be tuned by
choosing different numbers of buckets for the hash-map. 
We consider two scenarios along this dimension: \emph{low} contention, where the hash-map has 1000 buckets (hence, concurrent operations on the same bucket are rare); and \emph{high} contention, where the hash-map has only 10 buckets (frequent operations contend for a common bucket). 

We experiment each possible combination of transaction footprint with the two contention scenarios ($3 \times 2$ scenarios), running each combination up to 80 threads (10 cores running in no-SMT up to SMT-8 mode). 

%joao: revi o parágrafo a explicar os tipos de aborts, abaixo. Penso que ficou mais claro.
Figures \ref{plot:hashmap:largeRO:low} to \ref{plot:hashmap:small:high} present the throughput and discriminated abort rate of each experiment, averaged over five runs. 
Regarding the types of aborts, we distinguish transactional aborts, 
essentially caused by conflicting accesses to shared memory locations; non-transactional aborts, mostly caused by a locked SGL that kills ongoing transactions (only possible in HTM); and, of course, capacity aborts.
%On \name we group transactional and non-transactional aborts, since read-after-write transactional conflicts cause ROT non-transactional aborts. It is impossible to distinguish between these non-transactional aborts and those caused by SGL subscription.

%joao: chegaste a perceber o que é %aborts? :)
%nop..

As expected, the largest gains we observe are on large read-only transaction scenarios, with an impressive 576\% improvement of peak throughput on the low contention workload. This is the best-case scenario for \name, where most transactions are read-only, hence
run with no capacity bounds. This is in clear contrast with the prohibitive
capacity constraints of HTM; such capacity issues quickly escalate onto non-transactional aborts due to falling back to the SGL.

On the scenario with 50\% update transactions, where the majority of \name's transactions run as ROTs with a limited write set capacity, \name still proves to be the best approach on a low contention workload, with gains of up to 10\% peak throughput. Again, the fact that update transactions do not abort frequently for capacity reasons is the main reason. However, on the high contention workload, \name is not able to surpass regular HTM. Because of the quiescence phase, \name transactions take longer to abort (than in HTM), leading to a delay on the fall-back to the SGL.

On scenarios of small transactions, which mostly fall within \powerhtm's capacity bounds, \name is not able to surpass HTM. The added safety wait delays the execution of update transactions in \name, on both low and high contention workloads, a cost that is not compensated by relevant reductions on capacity aborts.

\begin{figure*}[t]
\centering
%\includegraphics[width=0.25\textwidth]{plots/tpcc-std-w40-throughput}
%\includegraphics[width=0.245\textwidth]{plots/tpcc-std-w40-aborts}
%\caption{TPC-C standard mix with low contention}
%\includegraphics[width=0.25\textwidth]{plots/tpcc-std-w4-throughput}
%\includegraphics[width=0.24\textwidth]{plots/tpcc-std-w4-aborts}
\includegraphics[width=0.49\textwidth]{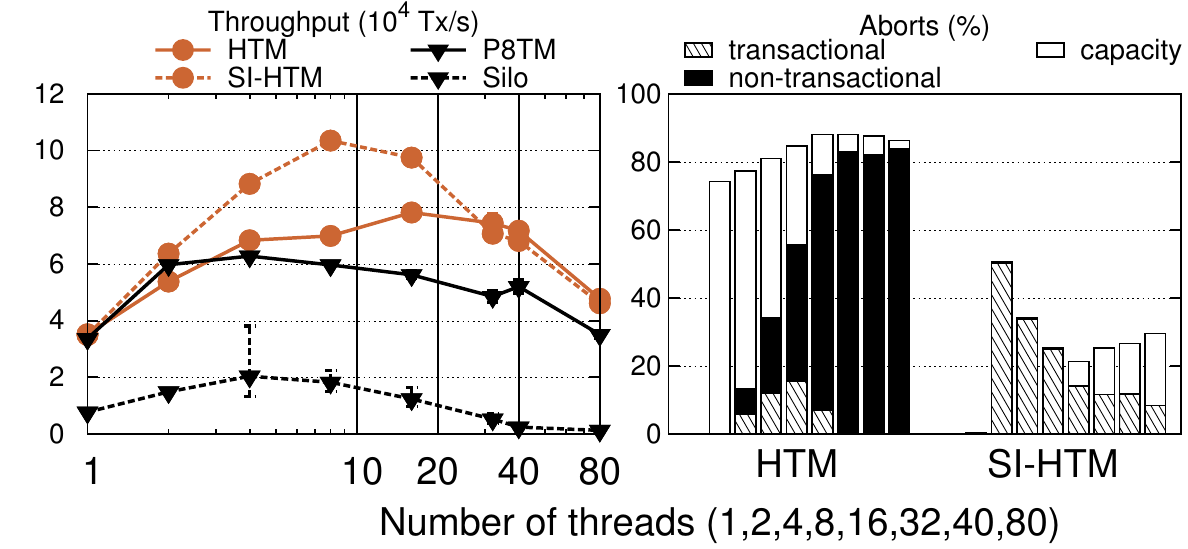}
\includegraphics[width=0.49\textwidth]{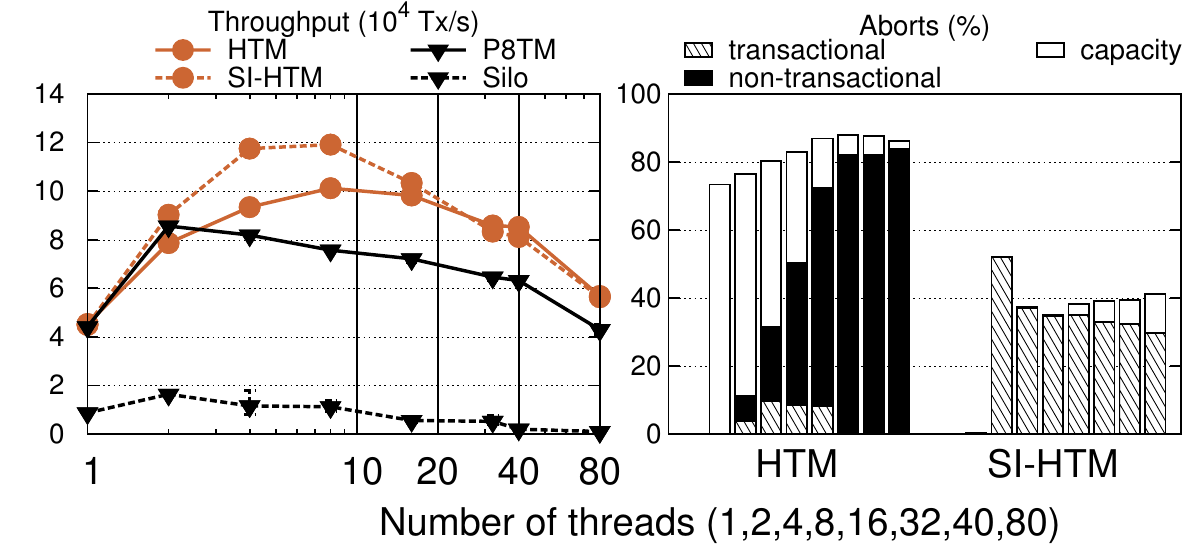}
\caption{TPC-C standard mix with low (left) and high (right) contention}
\label{plot:tpcc:std:low}
%\includegraphics[width=0.25\textwidth]{plots/tpcc-ro-w40-throughput}
%\includegraphics[width=0.245\textwidth]{plots/tpcc-ro-w40-aborts}
%\caption{TPC-C read dominated mix with low contention}
%\includegraphics[width=0.25\textwidth]{plots/tpcc-ro-w4-throughput}
%\includegraphics[width=0.24\textwidth]{plots/tpcc-ro-w4-aborts}
\includegraphics[width=0.49\textwidth]{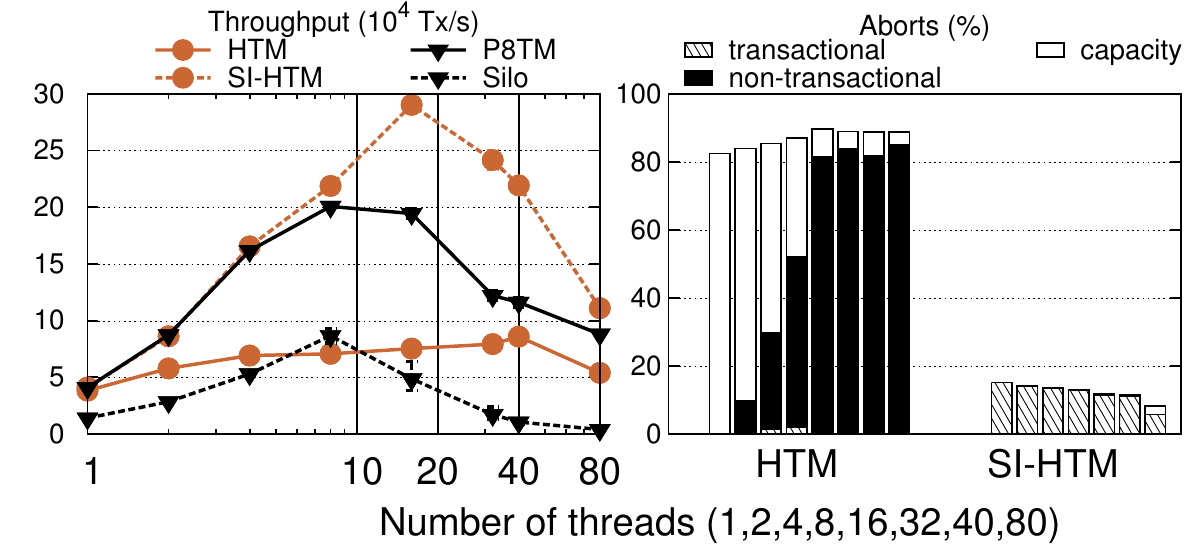}
\includegraphics[width=0.49\textwidth]{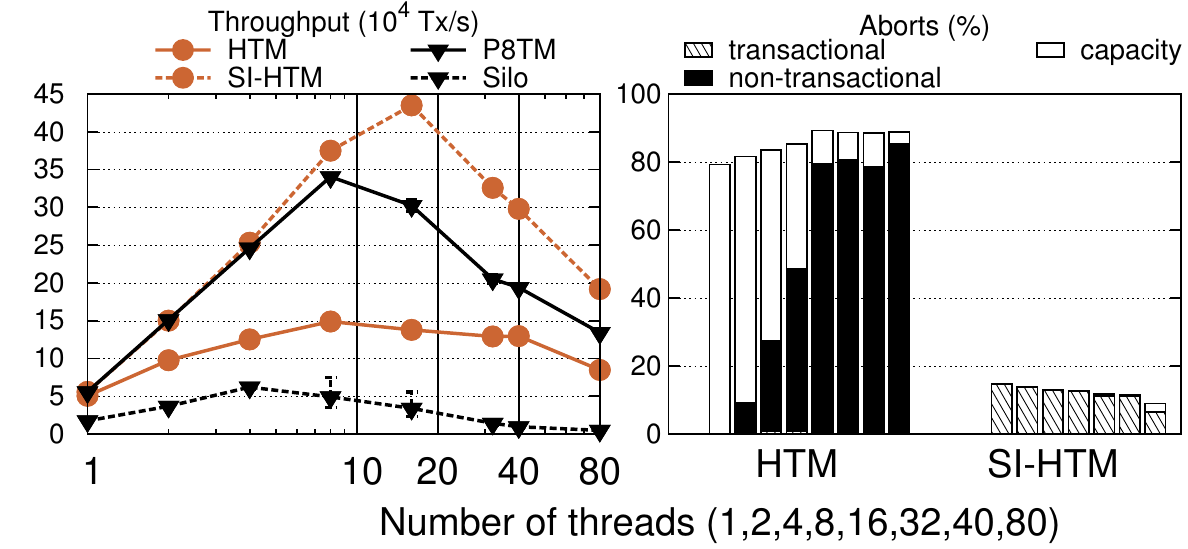}
\caption{TPC-C read dominated mix with low (left) and high (right) contention}
\label{plot:tpcc:read-dom:high}
\end{figure*}

HTM has been historically bad on SMT execution \cite{Diegues:2014:HTMLimits,Goel:2014:HTMPerf,Nakaike:2015:HTMComparison}, mostly due to sharing scarce hardware resources between SMT threads. 
Since a transaction's footprint in \name is limited only by its write-set, 
we expect that, in some workloads, multiple SMT transactions on the same core will finally fit in a shared TMCAM. In fact, we can observe on all low contention scenarios (even on short transactions) that \name behaves very well on SMT threads, scaling up to 32 threads, only showing signs of exhaustion at 40 threads, when the resources of each core start to be shared by more than four SMT threads. To the best of our knowledge, \name is the first HTM-based algorithm to consistently exhibit the power of SMT in low contention workloads.

\subsection{TPC-C benchmark}

To evaluate \name on a real-world application, we use the TPC-C benchmark, with a standard mix of transactions (i.e. \texttt{-s 4 -d 4 -o 4 -p 43 -r 45}), and a read-dominated mix (i.e. \texttt{-s 4 -d 4 -o 80 -p 4 -r 8}). The standard mix is primarily composed of update transactions, where roughly half of them have large transactional footprints. We also tested high and low-contention workloads of both mixes. Figures \ref{plot:tpcc:std:low} and \ref{plot:tpcc:read-dom:high} present the results for TPC-C under such configurations. We compare \name not only to HTM but also to P8TM \cite{Issa:2017:P8TM}, a HTM-based design for larger capacity transactions on \powerhtm (discussed in Section \ref{sec:relwork}); and Silo \cite{Tu:2013:STM:2517349.2522713}, a software-level optimistic concurrency control for in-memory databases. For a fair comparison, we disable record indexing in Silo, and the on-line adaptation of P8TM; this way, our analysis focuses exclusively on the core concurrency control performance of all solutions.

Overall, \name is able to improve peak performance of TPC-C's standard mix by 48\% on 8 threads, relatively to the best alternative (HTM). With 16 threads (SMT-2), the low contention workload still delivers very good results, albeit \name starts to show capacity issues, explained by TMCAM sharing, and exacerbated on higher thread counts. The improved resource usage of \name is especially evident on the read-dominated mix, where \name gracefully scales up to SMT-2 levels, improving 27\% in peak throughput over the best alternative (P8TM) and 300\% over base HTM. Still, the performance degrades on SMT-4 and SMT-8 modes, since much of a core's hardware is shared between the multiple SMT threads of that core.

\section{Related Work}
\label{sec:relwork}

%Since the introduction of HTM support in mainstream commercial processors by Intel and IBM, several experimental studies have aimed to characterize their performance and limitations \cite{Diegues:2014:HTMLimits,Goel:2014:HTMPerf,Nakaike:2015:HTMComparison}.
%These studies show that, although HTM excels with workloads that fit
%the hardware capacity, its performance and scalability are severely
%hampered even when a small percentage of transactions exceed the
%hardware's capacity.
%This is due to the need to execute such transactions using the sequential SGL
%fall-back mechanism, which prevents any form of parallelism.

Hybrid transactional memory (TM) designs \cite{Damron:2006:HyTM,Kumar:2006:HyTM,Dalessandro:2011:HNC,Matveev:2015:RHN} are the most visible effort that addresses the capacity issue in HTM-based programs. Hybrid TMs fall back to software-based TM (STM) implementations when transactions cannot successfully execute in hardware.
In contrast to \name, hybrid TMs do not change the capacity limits of the HTM; rather, they
aim at providing scalable STM fall-back paths.

%quiescence
Postponing a transaction's commit to a moment where the system state ensures
that committing the transaction will not result in correctness anomalies is not
a new concept of \name.
In another context, a notable example of a solution that relies on a similar
technique is the read-copy-update (RCU) synchronization mechanism  \cite{McKenneyRCU98}. Like \name, RCU allows multiple read-only threads to read directly from shared memory by having writer threads update a snapshot that is only committed at a later time when safety is ensured.
The RCU mechanism is entirely done in software and has been implemented
at user and kernel-level \cite{5386550}.
It requires the programmer to explicitly provide dedicated code in order to
create snapshots of the objects to update and ensure consistent pointers
to the right snapshot.
Read-Log-Update (RLU) extended RCU to allow a simpler programming model and
higher concurrency between readers and writers \cite{Matveev:2015:RLS:2815400.2815406}, by relying on techniques borrowed from the world of STM.

\sloppy
The works that are technically closer to \name are
HERWL~\cite{Felber:2016:HERWL}, SpRWL~\cite{Issa:2018:SpRWL} and P8TM~\cite{Issa:2017:P8TM}.
These works exploit the principle of making writers commit only when it is safe to do so.
%by relying on \powerhtm's advanced HTM features, namely the suspend-resume and ROT constructs.
However, their aim is clearly distinct than \name's, as they all offer strong consistency guarantees to programs based on read-write locks (HERWL and SpRWL) and in-memory transactions (P8TM).
One direct consequence of this key distinction is that none of them
relieve transactions from the cost of having the read sets of
update transactions tracked by the HTM (in HERWL and SpRWL) or by costly
software instrumentation of each read (in P8TM).
In contrast, since \name aims at weaker consistency guarantees, \name is able to completely free update transactions from read tracking. This fundamental advantage allows \name to clearly outperform P8TM for applications that are correct under SI, as Section \ref{sec:evaluation} shows.

In order to mitigate the worst-case scenarios where the quiescence phase yields
prohibitive latencies, P8TM proposes self-tuning techniques that revert to the baseline HTM support in unfavourable scenarios.
With minor adaptations, such self-tuning techniques may be incorporated in \name to improve its
performance in some of the unfavourable scenarios that Section \ref{sec:evaluation} identified.

An alternative direction to mitigate the constraints of
HTM capacity is by providing the weaker correctness guarantee of SI.
Two recent efforts have proposed new (or modified) hardware support
to implement SI on HTM.
Litz et al. \cite{Litz:2014:SRT:2541940.2541952} propose a multi-versioned memory architecture that implements SI for transactional programs running on parallel CPUs while Chen et al. \cite{Chen:2017:AGH:3079856.3080204}
pursue the same goal on GPUs. They propose a multi-versioned memory subsystem for transactional programs that run on a GPU, together with an
online method for eliminating the write skew anomaly on the fly.
When compared to \name, these systems allow higher concurrency degrees as they constitute \emph{full} SI implementations.
Further, the fact that they rely on a multi-versioned memory system obviates
the need for (and the cost of) \name's quiescence.
Nevertheless, \name is ready to use on commercially available systems,
which distinguishes it from these recent attempts to
combine SI and HTM.

On software, Litz et al. \cite{litz2016excite} implement SI by manipulating virtual memory mappings and using a copy-on-read mechanism with a customized page cache. Riegel et al. \cite{riegel2006snapshot} created an STM approach to SI by using a lazy multi-version mechanism. These approaches require instrumenting read and write operations of transactions, thus incumbering them over our HTM based implementation.
Litz et al. \cite{litz2015efficient} present a technique to automatically correct SI anomalies. Our work can also benefit from such a technique being used before deploying a new workload untested on SI systems.

IMDBs are among the domains where HTM capacity limits
constitute a major obstacle to adopt HTM to replace
current software-based concurrency control
schemes \cite{Leis2014ExploitingHT}.
Several previous proposals have leveraged concurrency control in IMDBs with HTM.
Leis et al. \cite{Leis2014ExploitingHT} use HTM transactions to run individual portions of a large transaction, with substantial code instrumentation.
Wang et al. \cite{wang2014using} leverage a software-based optimistic concurrency control mechanism with an optimized HTM-based commit stage.
Wu et al. \cite{wu2016scalable} adopt a similar HTM-assisted strategy, in this case using HTM transactions to perform optimized HTM-based pre-commit validation and writes to individual database records.
All these proposals use HTM as an auxiliary hardware mechanism to assist a software-based concurrency control.
In contrast, \name relies on HTM transactions -- more precisely, on ROTs -- as a first-class construct that runs full individual transactions, with non-instrumented memory accesses.

%Some recent proposals employ HTM as an auxiliary
%hardware mechanism to optimize software-based concurrency control approaches
%in IMDBs \cite{wang2014using,wu2016scalable}.
%For instance, Leis et al. \cite{Leis2014ExploitingHT} use HTM as a building block for executing each tuple access alone, while using timestamp ordering to commit the effects of all transactions together.
%Their proposal introduces substantial code instrumentation, both on writes and reads, in addition to the overheads of many transaction creations/commits per full transaction. \name supports large transactions without such overheads and complexity.

%Complementarily, the same paper also proposes a best-effort software tool
%that helps remove write skew anomalies.
%This tool can be used in conjunction with \name.

%Proposes modified hardware that supports SI on HTM, via a

%weak consistency (SI, etc) solutions

%\cite{Zamanian:2017:EMD:3055330.3055335}
%propose a scalable in-memory database system (NAM-DB) based on distributed transactions with SI coordinated using RDMA-enabled.

%\cite{Litz:2015:ECA:2695583.2693260}
%technique based on dynamic code and graph dependency analysis that automatically corrects existing snapshot isolation anomalies in transactional memory programs

%\cite{Fekete:2005:MSI:1071610.1071615}
%develop a theory that characterizes when nonserializable executions of applications can occur under SI, and propose how to modify the program logic of applications that are nonserializable under SI so that serializability will be guaranteed.

%\subsection{SI/Weak consistency for TM}

\section{Conclusions}
\label{sec:conclusions}

\name leverages the HTM features of the IBM POWER architectures with a software-based safety wait before commit to offer a restricted
implementation of Snapshot Isolation. 
As a main outcome, \name dramatically stretches the capacity bounds
of the underlying HTM with no hardware modifications, thus boosting the scalability of \powerhtm and opening it
to a much broader class of applications, like large-footprint transactions from the IMDB domain.

Our work emphasizes how important it is for commercially available HTM implementations to expose advanced HTM related mechanisms like ROTs and suspend-resume to the software layers. \name shows that such mechanisms can serve as building blocks to sophisticated software-hardware designs that enrich the baseline features.

%When compared to HTM-based concurrency control alternatives, \name exhibits unprecedented scalability, outperforming the pure HTM-based alternative by up to 47\% on TPC-C. %, while achieving speedups of up to 27\% relatively to reference software-based concurrency control solutions.

As future work, we plan to study advanced mechanisms to mitigate
the idle waiting time that RW transactions spend in \name. 
Among possible approaches, we envision a \emph{killing} alternative, where the group of already-completed transactions decides, based on system-efficient heuristics to kill the transactions that are taking too long to complete; and a \emph{batching} alternative, where
a completed transaction that predicts a long safety waiting uses such idle time to execute one (or more) subsequent transactions. We could also study how feasible a software based SI fallback path would be. %joao: discussao sobre a utilidade destes mecanismos avancados, future work (batching/killing, distribuicao - NAM)

\section{Acknowledgements}

Our thanks go to Pascal Felber for shepherding our paper, the anonymous reviewers who gave us valuable feedback, Brno University of Technology and the University of Neuchatel for providing us access to their IBM POWER8 machines.
This work is partially funded by FCT via projects UID/CEC/50021/2019 and PTDC/EEISCR/1743/2014.

\newpage
\clearpage

\section{Artifact Appendix}

%%%%%%%%%%%%%%%%%%%%%%%%%%%%%%%%%%%%%%%%%%%%%%%%%%%%%%%%%%%%%%%%%%%%%
 \subsection{Abstract}

Our artifact includes the algorithm described in the SI-HTM paper, the benchmarks used and scripts to reproduce the paper's results. There are no software dependencies to run our algorithm. The test machine should include an IBM POWER8 processor with at least 10 cores. Plotting scripts were included which produce the graphs presented in the paper. These graphs can be used to validate our results.

\subsection{Artifact check-list (meta-information)}

%{\small
 \begin{itemize}
   \item {\bf Algorithm:} yes, the main file of our algorithm is:\\
 POWER8TM/backends/p8tm-si/tm.h
   \item {\bf Program:} hashmap and TPCC, included
   \item {\bf Compilation:} GCC 5+
   \item {\bf Transformations:} no
   \item {\bf Binary:} no
   \item {\bf Data set:} none
   \item {\bf Run-time environment:} Linux
   %joao: in this hardware section, shoulnd't we be more generic and simply say IBM POWER8? same 
   \item {\bf Hardware:} IBM POWER8 processor
   %joao: maybe "no other processes sharing the same cores as SI-HTM"
   \item {\bf Run-time state:} no other processes sharing the same cores as SI-HTM
   %joao: "thread pinning" - not sure if this is clear; "couple of hours" - "a couple of hours", can't we be more precise?
   \item {\bf Execution:} specific thread pinning, two hours
   \item {\bf Metrics:} Execution time, number of operations, detailed specific abort counters
   \item {\bf Output:} graphs with throughput and abort rate
   \item {\bf Experiments:} compile the TinySTM back-end and run the given scripts
   \item {\bf How much disk space required (approximately)?:} 10 MBytes
   \item {\bf How much time is needed to prepare workflow (approximately)?:} 20 minutes
   \item {\bf How much time is needed to complete experiments (approximately)?:} 2 hours
   \item {\bf Publicly available?:} yes
   %joao: why "MIT"?
   \item {\bf Code/data licenses (if publicly available)?:} MIT
   \item {\bf Workflow frameworks used?:} None
   \item {\bf Archived?:}\\\url{https://doi.org/10.6084/m9.figshare.7378496}
 \end{itemize}

%%%%%%%%%%%%%%%%%%%%%%%%%%%%%%%%%%%%%%%%%%%%%%%%%%%%%%%%%%%%%%%%%%%%%
 \subsection{Description}

\subsubsection{How delivered}
 \url{https://doi.org/10.6084/m9.figshare.7378496}

10 MB

\subsubsection{Hardware dependencies}
 IBM POWER8 processor with 10+ cores

\subsubsection{Software dependencies}
 Gnuplot

\subsubsection{Data sets}
None

%%%%%%%%%%%%%%%%%%%%%%%%%%%%%%%%%%%%%%%%%%%%%%%%%%%%%%%%%%%%%%%%%%%%%
 \subsection{Installation}

tar -xvf artifact.tgz
 
cd POWER8TM/stms/tinystm;
make

%%%%%%%%%%%%%%%%%%%%%%%%%%%%%%%%%%%%%%%%%%%%%%%%%%%%%%%%%%%%%%%%%%%%%
 \subsection{Experiment workflow}
 
 cd si-htm
 
 - In each sub-folder (hashmap; tpcc) you will find scripts to run and plot the results of the respective benchmark

- Create a results folder for each of the benchmarks:

mkdir \textit{<benchmark>}/results

- Run each benchmark with the absolute path to the POWER8TM directory included in the artifact and the absolute path to the corresponding results directory:

bash run\_<benchmark>.sh \textit{<POWER8TM-dir> <results-dir>}

This creates a \textit{<results-dir>/date} sub-directory for that run

- Edit each benchmark's plot.sh script with the absolute path to gnuplot and eps driver

- Edit each benchmark's default-plots.sh script with the absolute path to your Gnuplot PostScript directory and the corresponding \textit{<results-dir/date>} directory

- Plot each benchmark: bash default-plots.sh

%%%%%%%%%%%%%%%%%%%%%%%%%%%%%%%%%%%%%%%%%%%%%%%%%%%%%%%%%%%%%%%%%%%%%
 \subsection{Evaluation and expected result}

The results of our artifact should be reproducible with small variations in single digit percentages. It will output graphs into the results/plots folder, which can be compared to those on the paper.
The raw data for the plots can be found in the results/summary folder, which include run time, number of transactions executed and number of aborts by type.

% %%%%%%%%%%%%%%%%%%%%%%%%%%%%%%%%%%%%%%%%%%%%%%%%%%%%%%%%%%%%%%%%%%%%%
% \subsection{Experiment customization}

% %%%%%%%%%%%%%%%%%%%%%%%%%%%%%%%%%%%%%%%%%%%%%%%%%%%%%%%%%%%%%%%%%%%%%
% \subsection{Notes}

% %%%%%%%%%%%%%%%%%%%%%%%%%%%%%%%%%%%%%%%%%%%%%%%%%%%%%%%%%%%%%%%%%%%%%
% \subsection{Methodology}

% Submission, reviewing and badging methodology:

% \begin{itemize}
%   \item \url{http://cTuning.org/ae/submission-20180713.html}
%   \item \url{http://cTuning.org/ae/reviewing-20180713.html}
%   \item \url{https://www.acm.org/publications/policies/artifact-review-badging}
% \end{itemize}

% \clearpage

\bibliographystyle{ACM-Reference-Format}
\bibliography{si-htm}
\end{document}